\documentclass[%
 reprint,
showpacs,
 amsmath,amssymb,
 aps,
prb,
]{revtex4-1}

\usepackage{graphicx}
\usepackage{dcolumn}
\usepackage{bm}
\usepackage[colorlinks = true, linkcolor = blue]{hyperref}
\usepackage[usenames,dvipsnames,svgnames,table]{xcolor}

\usepackage{tabularx}
\usepackage{dcolumn}
\begin{document}


\title{Spin-orbit coupling in methyl functionalized graphene}

\author{Klaus Zollner, Tobias Frank, Susanne Irmer, Martin Gmitra, Denis Kochan, and Jaroslav Fabian}
\affiliation{Institute for Theoretical Physics, University of Regensburg,
93040 Regensburg, Germany}

\date{\today}

\begin{abstract}
We present first-principles calculations of the electronic band structure and spin-orbit effects in graphene functionalized with methyl molecules in dense and dilute limits. The dense limit is represented by a $2 \times 2$ graphene supercell functionalized with one methyl admolecule. The calculated spin-orbit splittings are up to $0.6$~meV. The dilute limit is
deduced by investigating a large, $7 \times 7$, supercell with one methyl admolecule.
The electronic band structure of this supercell is fitted to a symmetry-derived effective
Hamiltonian, allowing us to extract specific hopping parameters including intrinsic, Rashba,
and pseudospin inversion asymmetry spin-orbit terms. These proximity-induced
spin-orbit parameters have magnitudes of about 1 meV, giant compared to pristine graphene whose intrinsic spin-orbit coupling is about 10 $\mu$eV. We find that
the origin of this giant local enhancement is the $sp^3$ corrugation and the breaking of local pseudospin inversion symmetry, as in the case of hydrogen adatoms. Similarly to hydrogen, also methyl acts as a resonant scatterer, with a narrow resonance peak near the charge neutrality point. We also calculate STM-like images showing the local charge densities at different energies around methyl on graphene.
\end{abstract}

\maketitle


\section{\label{sec:Intro}Introduction}
Spin-orbit coupling (SOC) effects in graphene~\cite{Han2014:NN} functionalized with adatoms and admolecules offer new possibilities for tailoring and manipulating electron spins, potentially leading to new spintronics devices~\cite{RevModPhys.76.323,fabian2007semiconductor}.
It has already been demonstrated that a giant enhancement of the rather weak intrinsic SOC of the Dirac electrons in graphene~\cite{Gmitra2009} can be achieved by adsorbates, such as light~\cite{PhysRevLett.103.026804,PhysRevLett.110.246602,Zhou20101405,
Irmer2014,PhysRevB.91.195408,PhysRevLett.109.186604,PhysRevLett.110.156602}
and heavy~\cite{Ma2012297,PhysRevB.82.125424,PhysRevLett.104.187201,PhysRevLett.109.266801,
PhysRevX.1.021001} adatoms. Inducing large SOC in graphene
is important for studying spin relaxation~\cite{Tuan2014:NP, Bundesmann2015:P} as
well as spin transport, in particular the spin-Hall effect
\cite{Balakrishnan2013, Ferreira2014:PRL}.
Here we show that giant SOC can be
also induced by organic molecules, taking methyl as their representative.\\
Methyl radical CH$_3$ is the most simple organic molecule.
Comprising one carbon bound to three hydrogen atoms, it forms an important building block for organic compounds.
It is a likely contaminant for graphene, especially in samples prepared by CVD (chemical vapor deposition) during which a H$_2$/CH$_4$ gas mixture is used~\cite{Bae2010}.
As a result, both H and CH$_3$ impurities could be expected.
Closer investigations revealed that hydrogen affects the thermal stability of CH$_3$ trapped on graphene~\cite{Berdiyorov2014,PhysRevB.82.165443}, forming clusters at high temperatures.\\
There have already been several investigations of methyl bonded to graphene,
including magnetic~\cite{Santos2012} and mechanical~\cite{Shenoy2010} effects. It has been shown
by density functional calculations, that a large class of organic molecules (including methyl) induce a spin-$\frac{1}{2}$ magnetic moment on graphene~\cite{Santos2012}, but in general, there is a strong dependence of the induced magnetism on the location, distribution and coverage of CH$_3$ on graphene.
Another reason to investigate methyl radicals is their similarity to hydrogen. As shown in Ref.~\cite{electronegativity}, an effective Pauling electronegativity of $2.28$ can be associated to CH$_3$, which almost coincides with that of hydrogen~\cite{Allen1989}, $2.20$.
Thus the bonding behavior of these adsorbates should be comparable. The important question is, will also the induced spin-orbit phenomena be similar?\\
In this paper we present first-principles calculations on methyl functionalized graphene in two different limits: dense and dilute.
For the dense limit we present the calculated electronic band structure and spin-orbit splittings of bands close to the Fermi level.
For the dilute limit we take a representative 7$\times$7 supercell with a single methyl admolecule, where we calculate the electronic band
structure, fit the bands at the Fermi level
to an effective symmetry-based Hamiltonian~\cite{PhysRevLett.110.246602,Irmer2014}, and
obtain the relevant SOC parameters: intrinsic, Rashba, and PIA (for pseudospin inversion asymmetry). Further, we investigate the nature of resonant scattering
of a methyl group bonded to graphene~\cite{Wehling2010}. Finally, we provide calculated scanning tunneling microscopy (STM) images and also study the magnetic moment formation by a single methyl radical.\\
Our main finding is that CH$_3$ admolecules induce a giant local SOC in
graphene, by a factor of 100 as compared to pristine
graphene's intrinsic SOC of about 10 $\mu$eV~\cite{Gmitra2009}. In the dilute limit, methyl acts as a resonant scatterer, with a resonance peak at $-8.8$~meV below the Dirac point, with a full width at half maximum (FWHM) of $4.9$~meV. Methyl prefers to bind in a configuration in which the hydrogen atoms point in the direction of the centers of the subjacent graphene honeycombs.
What would be the energetic cost of a methyl rotation on graphene? We have
calculated such activation energy and found that a rotation of the methyl group by $60^\circ$ around its bonding axis would require an energy of $0.17$~eV. Finally, we find, in agreement with previous studies~\cite{Santos2012} that CH$_3$ covalently bonded to graphene induces a spin-$\frac{1}{2}$  magnetic moment.\\
The paper is organized as follows. In Sec.~\ref{sec:Comp} we discuss our calculational
methods. Section~\ref{sec:Dense} presents the density functional theory (DFT) results of the electronic properties in the dense limit. Section~\ref{sec:Dilute} reports on the electronic structure and its phenomenological modeling for a $7 \times 7$ supercell with a single
methyl admolecule representing the dilute limit. Finally, in Sec.~\ref{sec:Charge} we discuss the induced electronic charge density, spin polarization, and STM images.
\section{\label{sec:Comp}Computational Methods}
Our first-principles calculations were carried out using the {\sc Quantum ESPRESSO}~\cite{QE-2009} suite based on density-functional theory~\cite{PhysRev.140.A1133} with plane waves and pseudopotentials~\cite{PhysRevB.43.1993, PhysRev.127.276}. 
We used fully relativistic projector augmented-wave~\cite{PhysRevB.50.17953} pseudopotentials with a Perdew-Burke-Ernzerhof~\cite{PhysRevLett.77.3865, PhysRevB.59.1758} exchange-correlation functional.
The kinetic energy cutoff for charge density and potential was $184$~Ry, the kinetic energy cutoff for wave functions was $46$~Ry, and the convergence threshold for self-consistency was $10^{-8}$~Ry.
In general, a $k$-point sampling of $10 \times 10 \times 1$ was used for self- and non-self-consistent calculations, except for the density of states where a higher sampling of $20\times20\times1$ was necessary and for the band structure along $\Gamma$-M-K-$\Gamma$ we used 80 discrete $k$-points.
We used a vacuum spacing of $15$~\AA~in the $z$ direction to simulate isolated graphene. Spin unpolarized ground states were used to study SOC effects. Structural relaxations were performed with the Broyden-Fletcher-Goldfarb-Shanno quasi-newton algorithm~\cite{quasinewton}.
\begin{figure}[htb]
\includegraphics[width=0.45\textwidth]{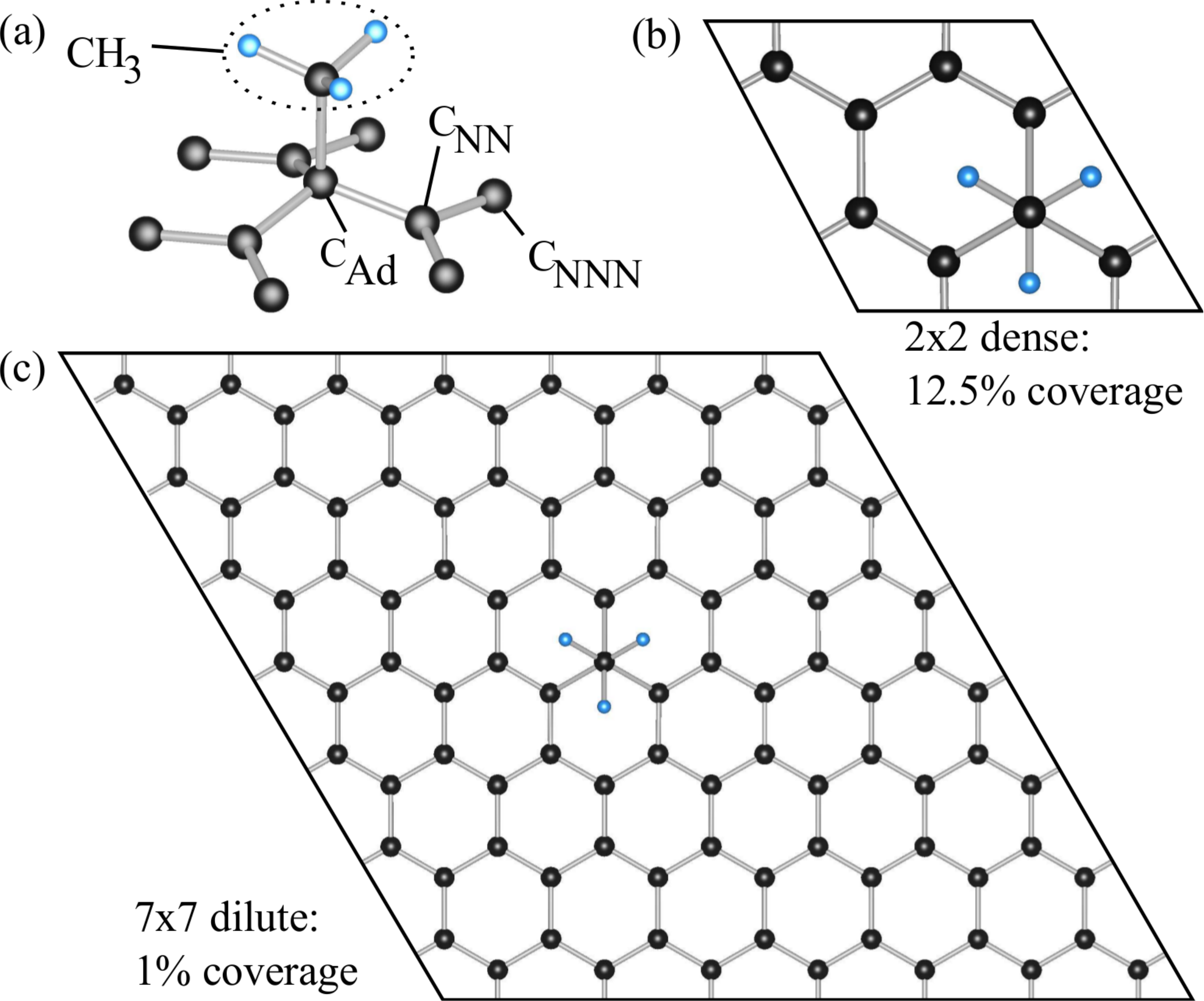}
\caption{\label{fig:unit_cells}(Color online) Structure of methyl functionalized graphene. (a) The geometric structure in the vicinity of the admolecule, with labels for the methyl group CH$_3$, the carbon atom that bonds the admolecule C$_{\textrm{Ad}}$, its three nearest, C$_{\textrm{NN}}$, and six next nearest, C$_{\textrm{NNN}}$, neighbors. (b) Unit cell of the dense (12.5\%) and (c) of the dilute (1\%) coverage limit.}
\end{figure}
In Fig.~\ref{fig:unit_cells}(a) we show the basis of the geometric structure used in our calculations, with atomic labels used throughout the paper. The methyl group is
labeled as CH$_3$, the carbon atom that bonds the admolecule as C$_{\textrm{Ad}}$, the nearest carbons as C$_{\textrm{NN}}$, and the next nearest as C$_{\textrm{NNN}}$. In equilibrium, the hydrogen atoms of CH$_3$ point towards the centers of subjacent graphene hexagons.
\section{\label{sec:Dense}Dense Limit}
The dense limit is represented by a 2$\times$2 supercell which is functionalized with a single methyl group (12.5\% coverage) [see Fig.~\ref{fig:unit_cells}(b)]. Structural relaxation shows, that for the 2$\times$2 supercell the carbon-admolecule C$_{\textrm{Ad}}$~-~CH$_{3}$ bond length is 1.607~\AA, the nearest-neighbor C$_{\textrm{Ad}}$~-~C$_{\textrm{NN}}$ bond length is 1.499~\AA, and the distance between the next-nearest-neighbors C$_{\textrm{NNN}}$~-~C$_{\textrm{NNN}}$ is 2.481~\AA~[see Fig.~\ref{fig:unit_cells}(a)].
The lattice constant $a$ is 2.479~\AA, somewhat greater than in pristine graphene (2.466~\AA). Similar to hydrogenated graphene, the chemisorption of the methyl group induces $sp^3$ hybridization. The carbon atom C$_{\textrm{Ad}}$, which hosts the methyl group, has an out of plane lattice distortion $\Delta$ of about 0.355~\AA. For comparison, in the hydrogenated graphene~\cite{PhysRevLett.110.246602}, the lattice distortion is about 0.36~\AA.\\
Bringing into contact an isolated methyl radical and a graphene sheet, the methyl radical starts to deform from a trigonal planar to a pyramidal configuration. A deformation energy of $\Delta E = 0.37$~eV is needed for such structural reconfiguration. In addition, forming
the covalent bond C$_{\textrm{Ad}}$ - CH$_{3}$ the graphene carbon atom C$_{\textrm{Ad}}$ experiences an out of plane distortion $\Delta$.
The bonding energy
\begin{equation}
E_{\textrm{B}} = -(E_{\textrm{Gr}+\textrm{CH}_3})+(E_{\textrm{Gr}}+E_{\textrm{CH}_3}),
\label{eq:binding}
\end{equation}
is defined as the difference between the ground-state energy of the methyl functionalized graphene $E_{\textrm{Gr}+\textrm{CH}_3}$, and the sum of energies of the deformed (pyramidally restructured) methyl group $E_{\textrm{CH}_3}$ and the locally corrugated graphene $E_{\textrm{Gr}}$ without the methyl group. Our calculated value is $E_{\textrm{B}}= 2.06$~eV.
\begin{figure}[htb]
\includegraphics[width=0.48\textwidth]{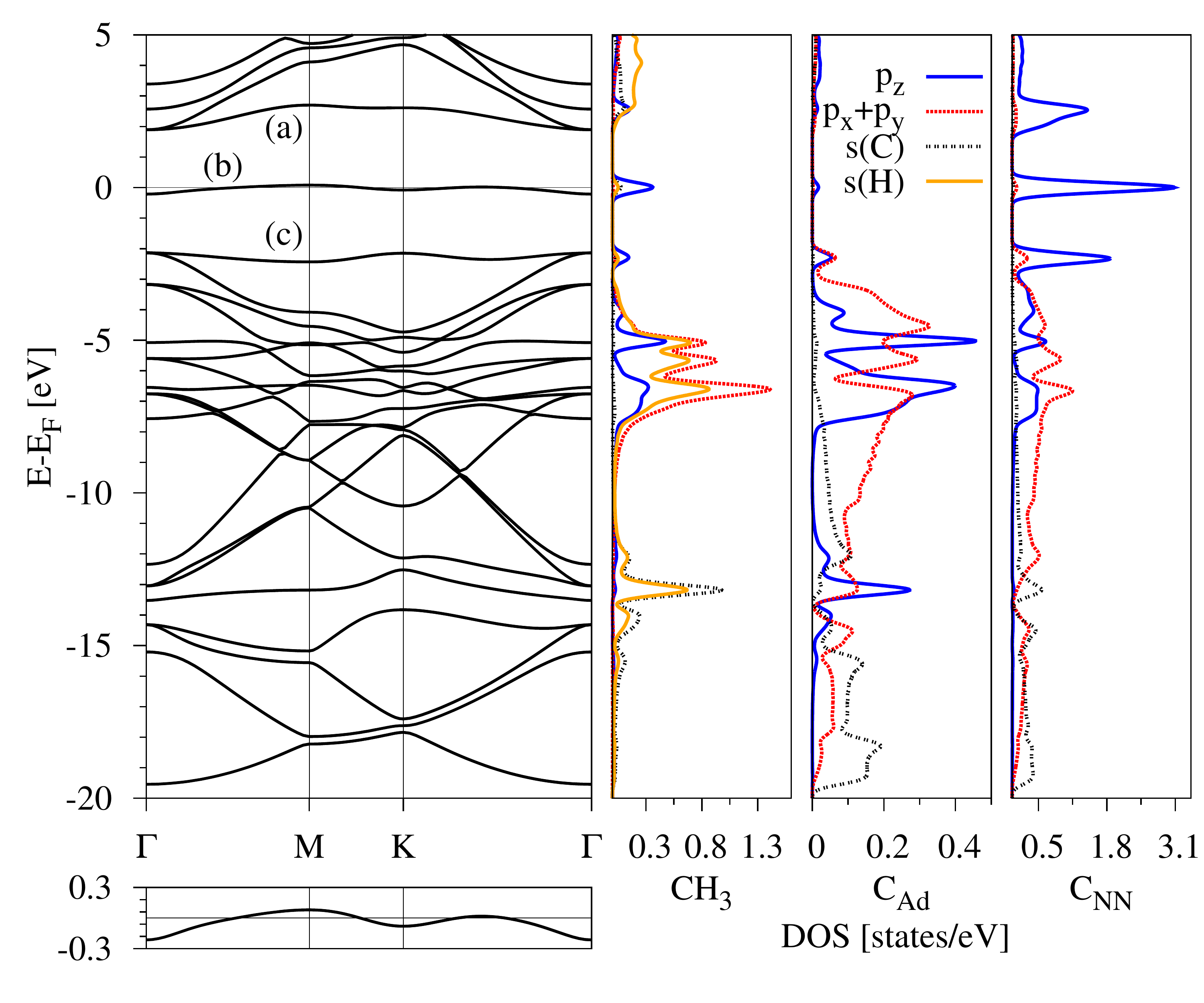}
\caption{\label{fig:bands_2x2}(Color online) Calculated band structure (left) of the methyl functionalized graphene for 2$\times$2 supercell configuration, with labels for the conduction (a), midgap (b), and valence (c) bands, respectively. The panel at the right shows the corresponding density of states for the admolecule and relevant carbon atoms. The contributions of different orbitals are indicated by the labeled lines. The panel below the band structure figure, shows a zoom on the midgap state in the energy region from $-0.3$~eV to $0.3$~eV.}
\end{figure}
\begin{figure}[htb]
\includegraphics[width=0.45\textwidth]{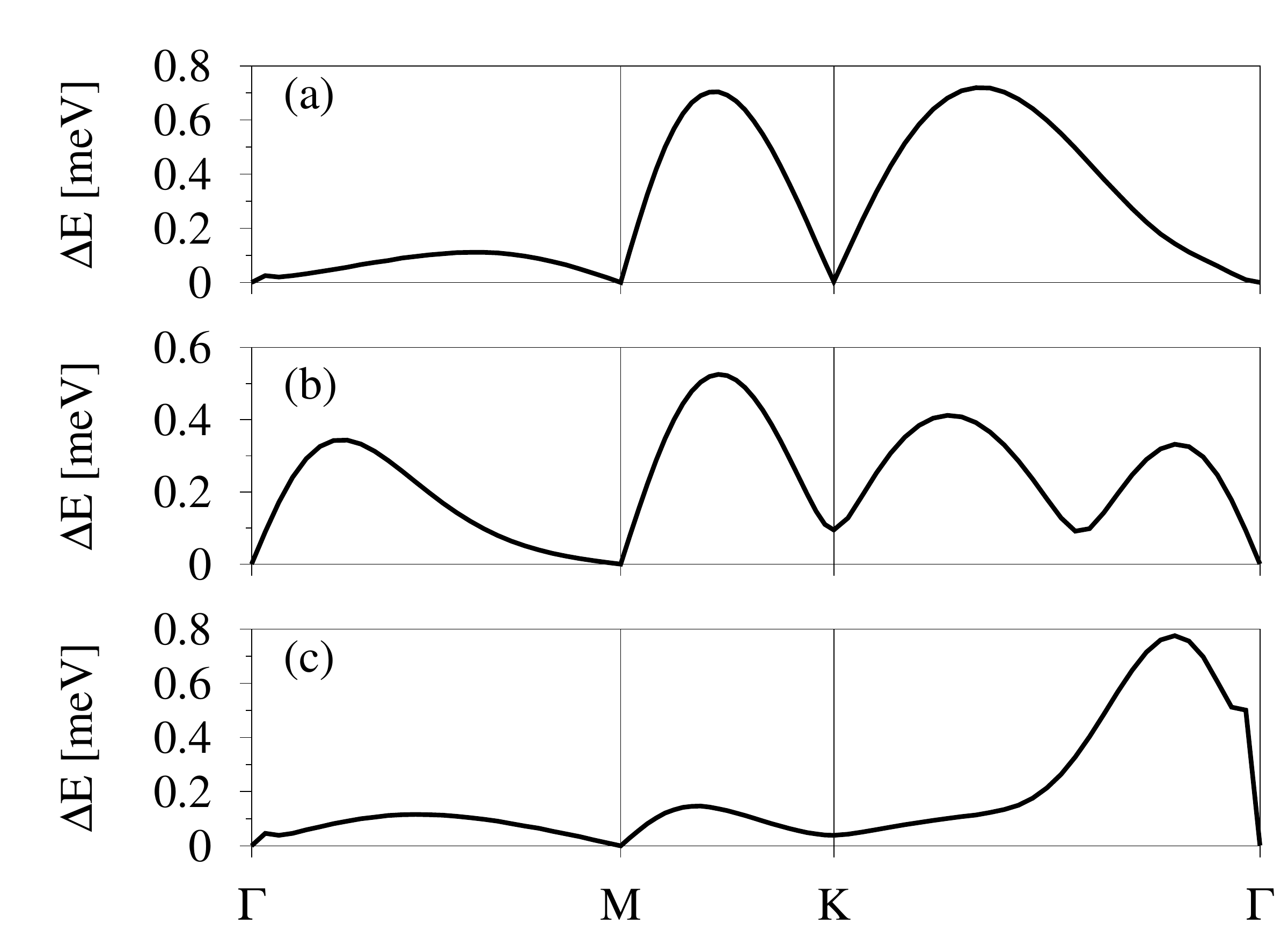}
\caption{\label{fig:splittings_2x2} First-principles computed SOC splittings for 2$\times$2 methyl functionalized graphene along the $\Gamma$-M-K-$\Gamma$ path: conduction (a), midgap (b), and valence (c) bands, respectively.}
\end{figure}
Figure~\ref{fig:bands_2x2} shows the first-principles computed band structure of the 2$\times$2 supercell. As in Refs.~\cite{PhysRevLett.110.246602, Irmer2014, Wehling2010}, a characteristic band appears at the Fermi level, which is induced by the methyl impurity. We will mainly focus on three bands, which we name conduction (a), midgap (b), and valence (c) bands, respectively.
The analysis of \textit{ab initio} data reveals that states near the Fermi level originate mainly from $p_z$ orbitals on the nearest neighbors C$_{\textrm{NN}}$. Bands in the energy windows ($-20$,$-15$)~eV and ($-12$,$-8$)~eV are mainly from $s$ and $p_x+p_y$ graphene carbon orbitals.
They correspond to the intact $\sigma$ bands of pristine graphene. In between, from $-15$~eV to $-12$~eV, there is a peak in the density of states (DOS)
which comes from $s$ and $p_z$ orbitals on CH$_{3}$ and C$_{\textrm{Ad}}$, respectively, which is a fingerprint of their covalent bonding.
The band structure at those energies is less dispersive reflecting its molecular states character.
The last DOS characteristic spans the energy window from $-8$~eV to $-3$~eV. There contributes mainly C$_{\textrm{Ad}}$ carbon with its $p$ orbitals
which provides an evidence of the internal hybridization towards $sp^3$ structure.
Moreover, at those energies also CH$_{3}$ admolecule shows its intrinsic character, namely the states participating in bondings among
the hydrogens' $s$ and carbon $p$ orbitals.\\
Figure~\ref{fig:splittings_2x2} shows SOC splittings of the three relevant bands, (a)-(c), near the Fermi level. The splitting maxima of the valence and conduction bands are nearly equal, while the midgap band is split less. However, all three are of the order of $0.6$~meV. At the time reversal points
$\Gamma$ and M the spin-orbit splittings are zero.
\section{\label{sec:Dilute}Dilute Limit}
\begin{figure}[htb]
\includegraphics[width=0.48\textwidth]{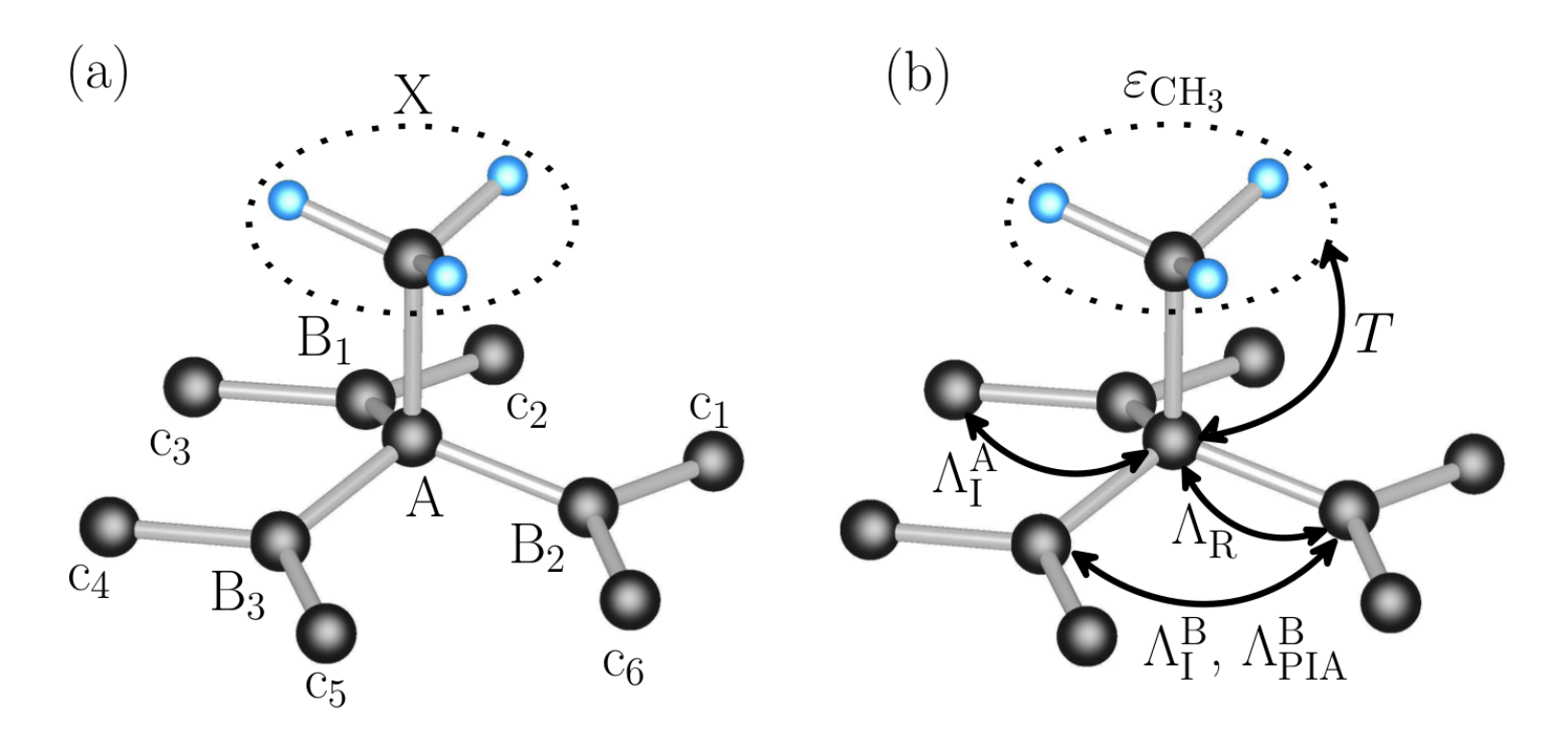}
\caption{\label{fig:hoppings}(Color online) Schematic plot showing notation and graphical representation of the minimal tight-binding Hamiltonian.
(a) Positions and labeling of the relevant atomic sites whose $p_z$ orbitals enter the model Hamiltonian. Shown are: X $={\textrm{CH}_3}$, A = C$_{\textrm{Ad}}$, three nearest, B$_i$, $i= 1,2,3$, and six next-nearest, c$_j$, $j=1,\ldots,6$, neighbors, respectively.
(b) Sketch of the dominant orbital and spin-orbital hoppings near the admolecule, all carbons in graphene lattice are coupled by the nearest-neighbor hopping
$t=2.6$\,eV (not shown).}
\end{figure}

\begin{figure}[htb]
\includegraphics[width=0.48\textwidth]{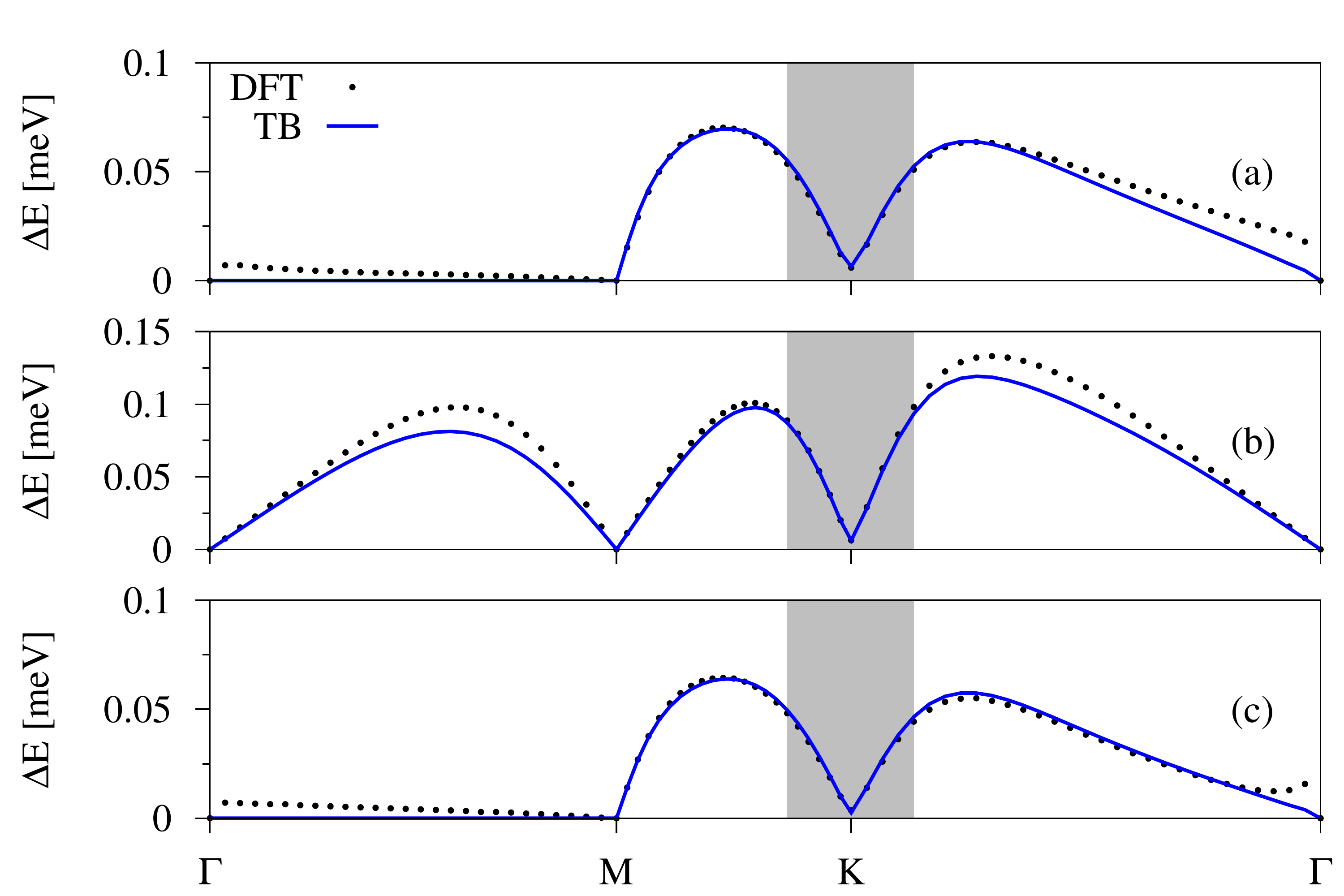}
\caption{\label{fig:splittings_7x7}(Color online) Calculated spin-orbit splittings along the $\Gamma$-M-K-$\Gamma$ path for the conduction (a), midgap (b), and valence band (c), respectively. First-principles data (dotted) are well reproduced by the tight-binding model (solid) with Hamiltonian $\mathcal{H}_{\textrm{orb}}+\mathcal{H}_{\textrm{so}}$, Eqs.~(\ref{eq:Horb}) and (\ref{eq:Hso}), using $T = 7.6$~eV, $\varepsilon_{\textrm{CH}_3} = -0.19$~eV, $\Lambda_{\textrm{I}}^{\textrm{A}} = -0.77$~meV, $\Lambda_{\textrm{I}}^{\textrm{B}} = 0.15$~meV, $\Lambda_{\textrm{R}}^{} = 1.02$~meV, and $\Lambda_{\textrm{PIA}}^{\textrm{B}} = -0.69$~meV. The least square fitting was performed in the shaded regions around K.}
\end{figure}

The dilute limit is represented by a functionalized 7$\times$7 supercell comprising a single methyl group (1\% coverage) [see Fig.~\ref{fig:unit_cells}(c)]. Structural relaxation shows that for the 7$\times$7 supercell the carbon-admolecule C$_{\textrm{Ad}}$ - CH$_{3}$ bond length equals 1.583~\AA, the nearest-neighbor C$_{\textrm{Ad}}$ - C$_{\textrm{NN}}$ bond length equals 1.509~\AA and the next-nearest-neighbor distance C$_{\textrm{NNN}}$ - C$_{\textrm{NNN}}$ equals 2.464~\AA. The carbon atom C$_{\textrm{Ad}}$ has an out of plane lattice distortion $\Delta$ of about 0.410~\AA, which is more than in the dense limit. The reason is the more relaxed $\sigma$ bond network in the dilute limit which allows a more ideal $sp^3$ tetrahedral
distortion.
In both cases, there is a competition between the C$_{\textrm{Ad}}$ - CH$_{3}$ and C$_{\textrm{Ad}}$ - C$_{\textrm{NN}}$ bonds in forming the 109.5$^\circ$ tetrahedral angle. This competition tends to modify $\Delta$ and also the in-plane alignment of the C$_{\textrm{NN}}$ - C$_{\textrm{NNN}}$ bonds.
In the dilute limit the in-plane alignment is geometrically not constrained so tightly as in the dense limit case, what results in a larger $\Delta$
and shorter C$_{\textrm{NNN}}$ - C$_{\textrm{NNN}}$ bond distance. This behavior is similar, e.g. with hydrogenated or fluorinated graphene~\citep{Irmer2014,PhysRevLett.110.246602}.
The bonding energy in the dilute limit is found to be $2.46$~eV, which is larger than in the dense case. However, the magnitude is very close to the one for hydrogen $E_{\textrm{B}} = 2.9$~eV~\cite{PhysRevLett.110.246602}, supporting the fact of a covalent bonding.\\
Before we discuss the \textit{ab initio} results let us introduce our minimal tight-binding Hamiltonian.
For the description of the orbital part, we employ a nearest neighbor tight-binding Hamiltonian~\cite{Wehling2010, Robinson2008} based on carbon $p_z$ orbitals since those are mainly contributing to states around the Fermi level.
Hamiltonian $\mathcal{H}_{\textrm{orb}}$ consists of an on-site energy $\varepsilon_{\textrm{CH}_3}$ term for the methyl group, a hybridization $T$ for the hopping between the adsorbate and host graphene carbon and the standard nearest neighbor hopping $t = 2.6$~eV for the remaining carbons in the lattice.
For simplicity, we model the methyl group as a single energy level with one effective $p_z$ orbital that bonds on top of a carbon atom. The orbital Hamiltonian reads as
\begin{eqnarray}
\mathcal{H}_{\textrm{orb}} &&=  \varepsilon_{\textrm{CH}_3}\sum_{\sigma}X_{\sigma}^{\dagger}X_{\sigma}+
T\sum_{\sigma}(X_{\sigma}^{\dagger}A_{\sigma}+A_{\sigma}^{\dagger}X_{\sigma})\nonumber\\
&&-t\sum_{B_{j}\in \textrm{C}_{\textrm{NN}}}\sum_{\sigma}(A_{\sigma}^{\dagger}B_{j,\sigma}+B_{j,\sigma}^{\dagger}A_{\sigma})\nonumber\\
&&-t\sum_{\langle i,j\rangle}\sum_{\sigma}(B_{i,\sigma}^{\dagger}c_{j,\sigma}+c_{j,\sigma}^{\dagger}B_{i,\sigma})\nonumber\\
&&-t\sum_{\langle i,j\rangle}\sum_{\sigma}(c_{i,\sigma}^{\dagger}c_{j,\sigma}+c_{j,\sigma}^{\dagger}c_{i,\sigma}),
\label{eq:Horb}
\end{eqnarray}
where $\langle i,j \rangle$ denotes the summation over the nearest neighbors. The operator $X_{\sigma}^{\dagger} ~(X_{\sigma})$ creates (annihilates) an electron with spin $\sigma$ in the effective $p_z$ orbital on the methyl group. Similarly,
$c_{i,\sigma}^{\dagger} ~ (c_{j,\sigma})$ are the creation (annihilation) operators for $p_z$ orbitals of graphene carbon atoms. Specifically, we introduce $A_{\sigma}^{\dagger}~(A_{\sigma})$ and $B_{i\sigma}^{\dagger}$ ($B_{i\sigma}^{}$) as the creation (annihilation) operators for C$_{\textrm{Ad}}$ carbon (assuming it is on sublattice A) and its three nearest neighbors, C$_{\textrm{NN}}$ (on sublattice B), respectively. The notation and labeling are illustrated in Fig.~\ref{fig:hoppings}.
\begin{figure}[htb]
\includegraphics[width=0.48\textwidth]{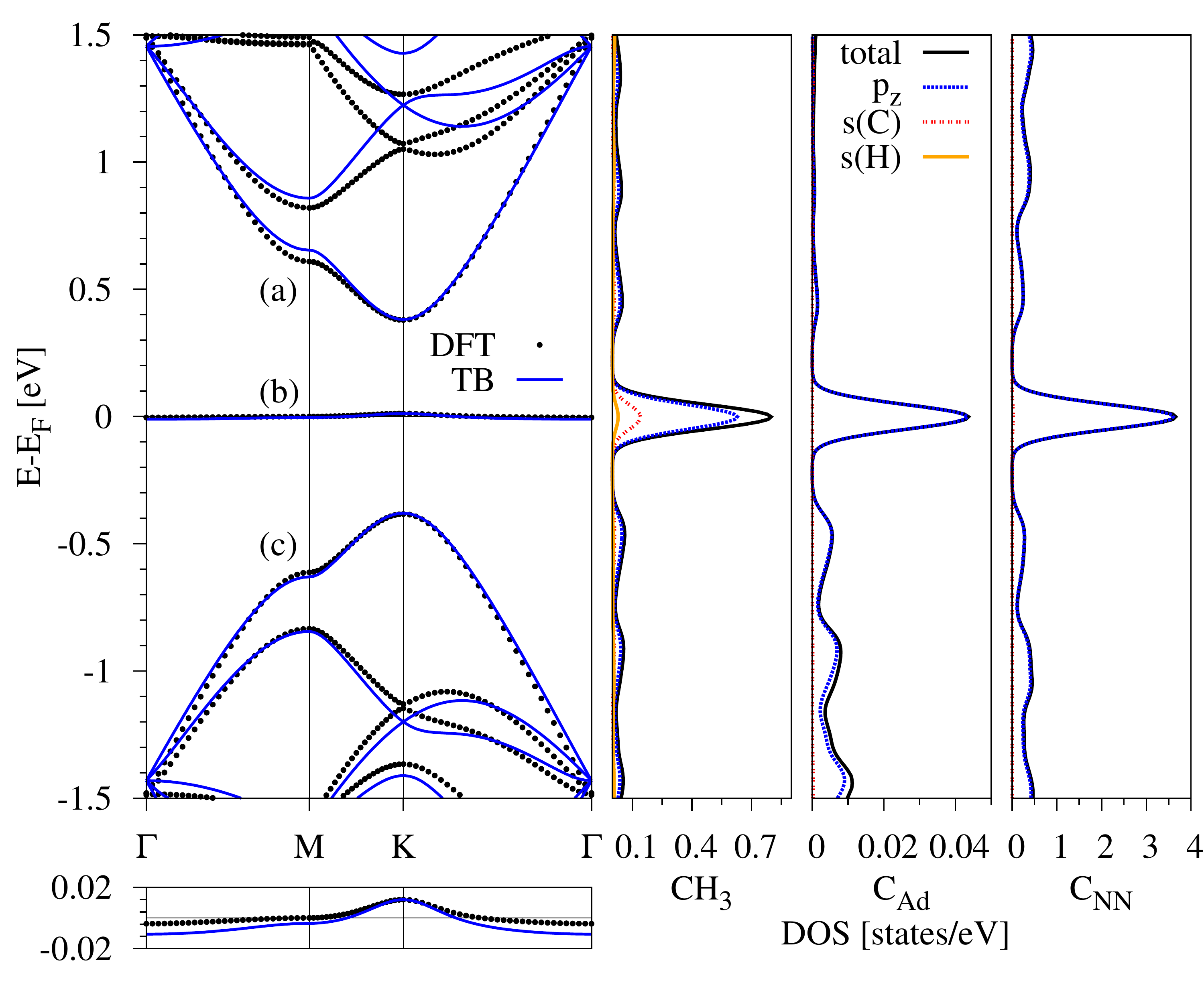}
\caption{\label{fig:bands_7x7}(Color online) Calculated electronic band structure of the methyl functionalized graphene in the dilute limit represented
by $7 \times 7$ supercell. Left panel: First-principles (dotted black) band structure along with the tight-binding fit (solid blue) for the conduction (a), midgap (b) and valence band (c), respectively. Right panel: The corresponding broadened orbital resolved density of states for the admolecule and atoms in its vicinity, different orbital contributions are indicated by the labeled lines. Panel below the band structure figure shows a zoom on the midgap band including the tight-binding fit in the energy region from $-0.02$~eV to $0.02$~eV.}
\end{figure}
Figure~\ref{fig:bands_7x7} shows the DFT calculated spin-unpolarized electronic band structure of the fully relaxed $7 \times 7$ supercell (dotted lines) along with the tight-binding fits (solid lines).
The spectrum in the vicinity of the Fermi level shows three characteristic bands, which, in analogy with the dense limit, we call conduction (a), midgap (b) and, valence (c) bands, respectively.
These bands can be fitted by two parameters $T = 7.6$~eV and $\varepsilon_{\textrm{CH}_3} = -0.19$~eV that enter the orbital Hamiltonian $\mathcal{H}_{\textrm{orb}}$ in Eq.~(\ref{eq:Horb}).
They were obtained by minimizing the least-square differences between the first-principles and the tight-binding computed band structures considering the
three bands around the Fermi level. The shaded regions around the K point in Fig.~\ref{fig:splittings_7x7} show the k-space range employed in the fitting.\\
The midgap state becomes more localized, as a consequence of flatter band dispersion when compared to the dense limit. This indicates a weaker interaction between the supercell periodic images unlike to the case of dense functionalization limit.~The energy bandwidth over which the midgap band extends is here only $10$~meV, whereas in the dense limit it is $300$~meV.
The main contributions to the three relevant bands (a)--(c) come from $p_z$ orbitals on ${\textrm{CH}_3}$ and the nearest-neighbor carbon atoms C$_{\textrm{NN}}$; see the orbital resolved density of states in the right panel of Fig.~\ref{fig:bands_7x7}.
This fully acknowledges our minimal tight-binding Hamiltonian model which implements only $p_z$ orbitals.\\
To describe SOC effects, we extract the SOC parameters from our \textit{ab initio} data by employing a minimal spin-orbit coupling Hamiltonian $\mathcal{H}_{\textrm{so}}$~\cite{Irmer2014,PhysRevLett.110.246602,Bundesmann2015:P}:\\
\begin{eqnarray}
\mathcal{H}_{\textrm{so}}&& =
\frac{\textrm{i}\Lambda_{\textrm{I}}^{\textrm{A}}}{3\sqrt{3}}\sum_{c_{j}\in \textrm{C}_{\textrm{NNN}}}\sum_{\sigma}\left[A_{\sigma}^{\dagger}\nu_{ij}(\hat{s}_{z})_{\sigma\sigma}c_{j,\sigma}+\textrm{H.c.}\right]\nonumber\\
&&
+\frac{\textrm{i}\Lambda_{\textrm{I}}^{\textrm{B}}}{3\sqrt{3}}\sum_{\langle\langle i,j\rangle\rangle}\sum_{\sigma}B_{i,\sigma}^{\dagger}\nu_{ij}(\hat{s}_{z})_{\sigma\sigma}B_{j,\sigma}\nonumber\\
&&
+\frac{2\textrm{i}\Lambda_{\textrm{R}}^{\textrm{}}}{3}\sum_{B_{j}\in \textrm{C}_{\textrm{NN}}}\sum_{\sigma\neq\sigma'}\left[A_{\sigma}^{\dagger}(\hat{\bm{s}}\times\bm{d}_{Aj})_{z,\sigma\sigma'}B_{j,\sigma'}+\textrm{H.c.}\right]\nonumber\\
&&
+\frac{2\textrm{i}\Lambda_{\textrm{PIA}}^{\textrm{B}}}{3}\sum_{\langle\langle i,j\rangle\rangle}\sum_{\sigma\neq\sigma'}B_{i,\sigma}^{\dagger} (\hat{\bm{s}}\times\bm{D}_{ij})_{z,\sigma\sigma'}B_{j,\sigma'}\nonumber\\
&&
+\frac{\textrm{i}\lambda_{\textrm{I}}^{\textrm{}}}{3\sqrt{3}}\mathop{\sum\nolimits'}_{\langle\langle i,j\rangle\rangle}\sum_{\sigma}c_{i,\sigma}^{\dagger}\nu_{ij}(\hat{s}_{z})_{\sigma\sigma}c_{j,\sigma}.
\label{eq:Hso}
\end{eqnarray}
Here, symbol $\hat{\bm{s}}$ represents the array of Pauli matrices. The sign factor $\nu_{ij}$ equals $-1$ ($+1$) for a (counter-) clockwise hopping path connecting next-nearest neighbors.
Vectors $\bm{d}_{ij}$ and $\bm{D}_{ij}$ are unit vectors in the $xy$-plane, pointing from site $j$ to $i$.
The last term in Eq.~(\ref{eq:Hso}) is the global intrinsic SOC of graphene with $\lambda_{\textrm{I}} = 12~\mu\textrm{eV}$\cite{Gmitra2009}.
The primed sum therein runs over the sites that are not coupled by $\Lambda_{\textrm{I}}^{\textrm{A}} $ nor $\Lambda_{\textrm{I}}^{\textrm{B}} $.
The fact that the orbital and spin-orbital energy scales are different by three orders of magnitude, allows us to fit the orbital Hamiltonian $\mathcal{H}_{\textrm{orb}}$, Eq.~(\ref{eq:Horb}), ignoring any SOC contributions.\\
The spin-orbit splittings along the high symmetry path $\Gamma$-M-K-$\Gamma$ within the first Brillouin zone for the three bands around the Fermi level are shown  in Fig.~\ref{fig:splittings_7x7}.
The splittings vanish at the time reversal points $\Gamma$ and M. The maxima of the splittings for the conduction (a), midgap (b), and valence band (c) are of the order of $0.1$~meV, where the largest SOC splitting is experienced by the midgap band, which is in contrast to the dense limit case.
The multiband least-square fits were performed in the vicinity of the K point and we extract the following SOC-parameters: $\Lambda_{\textrm{I}}^{\textrm{A}} = -0.77$~meV, $\Lambda_{\textrm{I}}^{\textrm{B}} = 0.15$~meV, $\Lambda_{\textrm{R}}^{} = 1.02$~meV and $\Lambda_{\textrm{PIA}}^{\textrm{B}} = -0.69$~meV.
We observed that the main shape of the spin-orbit splitting curves is reproduced only by $\Lambda_{\textrm{PIA}}^{\textrm{B}}$.
Thus, similar to hydrogenated graphene~\cite{PhysRevLett.110.246602}, effects of SOC originate mainly from the breaking of local pseudospin inversion symmetry. SOC parameters for the methyl functionalized graphene are in magnitude comparable with its hydrogenated counterpart; for comparison, see Table \ref{tab:comparison}.\\
We stress that only spin-orbit {\it couplings} as obtained from the fitting of DFT to tight-binding model can be meaningfully compared with the corresponding parameters in pristine graphene. The spin-orbit {\it splittings} depend on the admolecule concentration, and by themselves are of little use when compared with
the pristine graphene or with graphene functionalized by different adsorbates. The CH$_3$ induced SOC parameters, $\Lambda_{\textrm{I}}$, $\Lambda_{\textrm{R}}$, $\Lambda_{\textrm{PIA}}$, are of the order of 1~meV --- 100-times larger than the intrinsic SOC parameter $\lambda_{\textrm{I}}\simeq 10 \mu$eV characterizing the unperturbed graphene.\\
Additionally to the 7$\times$7 supercell configuration, we also calculated a 5$\times$5 supercell structure. The first-principles data for this case can also be nicely fitted with our tight-binding model, however, with slightly modified orbital and SOC parameters: $T = 7.6$~eV, $\varepsilon_{\textrm{CH}_3} = -0.16$~eV, $\Lambda_{\textrm{I}}^{\textrm{A}} = -0.39$~meV, $\Lambda_{\textrm{I}}^{\textrm{B}} = 0.095$~meV, $\Lambda_{\textrm{R}}^{} = 1.01$~meV and $\Lambda_{\textrm{PIA}}^{\textrm{B}} = -0.71$~meV.
The fact that the values for both supercells are similar, see Table~\ref{tab:comparison}, confirms that our model is robust and reliable for the dilute methyl functionalized graphene. Our orbital results are in agreement with a $4\times4$ supercell calculation already reported in Ref.~\cite{Wehling2010}.\\
\begin{table*}[htbp]
\caption{\label{tab:comparison}
Orbital and spin-orbital tight-binding parameters which fit the band structure for methyl functionalized graphene for 5$\times$5 and 7$\times$7 supercells, respectively. Since the different supercell values are comparable, the robustness of the proposed tight-binding model is well acknowledged. Comparison of the fitted parameters with hydrogenated and fluorinated graphene shows certain similarity between CH$_{3}$ and H graphene functionalization.\\
}
\begin{ruledtabular}
\squeezetable
\begin{tabular*}{0.3\textwidth}{lllccccl}
X (Adsorbate)&
\textrm{n$\times$n}&
\textrm{$T$[eV]}&
\textrm{$\varepsilon_{\textrm{X}}$(eV)}&
\textrm{$\Lambda_{\textrm{I}}^{\textrm{A}}$(meV)}&
\textrm{$\Lambda_{\textrm{I}}^{\textrm{B}}$(meV)}&
\textrm{$\Lambda_{\textrm{PIA}}^{\textrm{B}}$(meV)}&
\textrm{$\Lambda_{\textrm{R}}^{\textrm{}}$(meV)}\\
\colrule
\\
CH$_3$&5$\times$5 & $7.6$ & $-0.16$ & $-0.39$ & $0.095$ & $-0.71$ & $1.01$\\
&7$\times$7 & $7.6$ & $-0.19$ & $-0.77$ & $0.15$ & $-0.69$ & $1.02$\\
\\
H\footnote{taken from Ref.~\cite{PhysRevLett.110.246602}}&5$\times$5 & $7.5$ & ~$0.16$ & $-0.21$ & - & $-0.77$ & $0.33$\\
\\
F\footnote{taken from Ref.~\cite{Irmer2014}}&7$\times$7 & $6.1$ & $-3.3$ & -& $3.2$ & $7.9$ & $11.3$\\
& 10$\times$10& $5.5$ & $-2.2$ & - & $3.3$ & $7.3$ & $11.2$\\
\end{tabular*}
\end{ruledtabular}
\end{table*}
\begin{figure}[htb]
\includegraphics[width=0.49\textwidth]{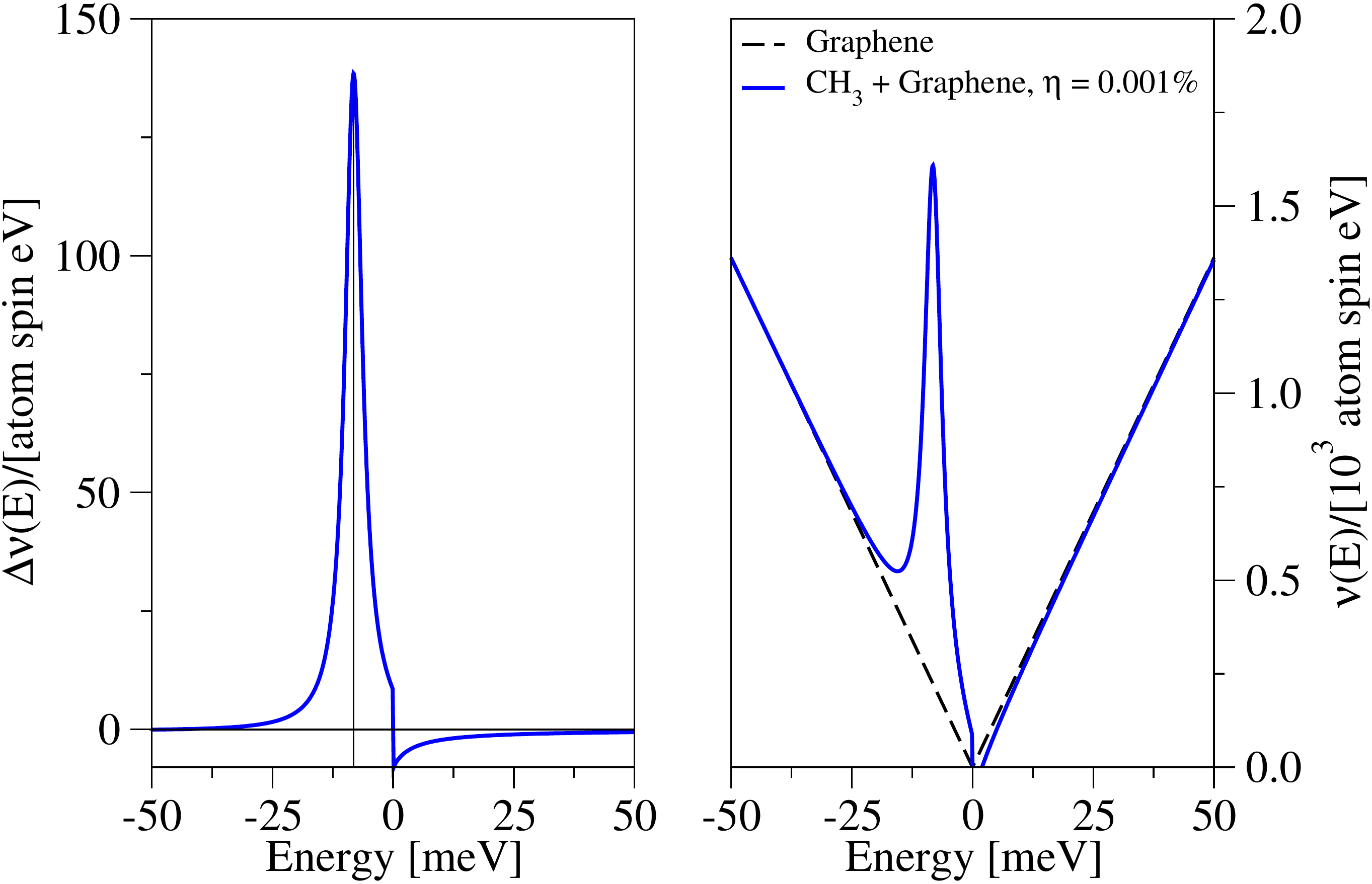}
\caption{\label{fig:resonant_scatter}(Color online) Left panel: change in DOS $\Delta\nu$, Eq.~(\ref{change_DOS}), for a single impurity limit with parameters $T=7.6$~eV and $\varepsilon_{\textrm{CH}_3} = -0.19$~eV. The resonance peak appears at $E\simeq -8.8$ meV with a $\textrm{FWHM}\simeq 4.9$~meV. Right panel: perturbed DOS, $\nu(E) = \nu_0(E)+\eta\Delta\nu(E)$, for the admolecule concentration $\eta = 0.001$\% (solid line) and the unperturbed pristine graphene DOS (dashed line) near the charge neutrality point.}
\end{figure}
Similarity between hydrogenated~\cite{PhysRevLett.110.246602} and methyl functionalized graphene indicates that the latter should also act as a resonant scatterer~\cite{Wehling2010}.
To describe the single admolecule limit, we downfold the tight-binding Hamiltonian $\mathcal{H}_{\textrm{orb}}$, Eq.~(\ref{eq:Horb}), by removing the admolecule $p_z$ orbital obtaining~\cite{PhysRevLett.112.116602,Irmer2014,Bundesmann2015:P,NEW}
\begin{equation}
\mathcal{H}'_{\textrm{fold}}(E)=\sum_{\sigma}\alpha(E)A_{\sigma}^{\dagger}A_{\sigma}^{}
\end{equation}
with
\begin{equation}
\alpha(E) = \frac{T^2}{E-\varepsilon_{\textrm{CH}_3}}.
\end{equation}
The change in the DOS, $\Delta\nu(E)$, due to a single methyl admolecule, is then given by
\begin{equation}
\Delta\nu(E)=\frac{1}{\pi}\textrm{Im}\left[\frac{\alpha(E)}{1-\alpha(E)G_0(E)}\frac{\partial}{\partial E}G_0(E)\right],
\label{change_DOS}
\end{equation}
where $G_0(E)$ is the Green's function per atom and spin for the unperturbed pristine graphene
\begin{equation}
G_0(E) \simeq \frac{E}{D^2}\,\ln\left|\frac{E^2}{D^2-E^2}\right|-\textrm{i}\pi\frac{|E|}{D^2}\,\Theta(D-|E|)
\end{equation}
with the effective graphene bandwidth $D = \sqrt{\sqrt{3}\pi}t\simeq 6$\,eV; for details see Refs.~\cite{Wehling2010, PhysRevLett.112.116602,Bundesmann2015:P,NEW}.\\
Employing our best-fit orbital tight-binding parameters $T=7.6$~eV and $\varepsilon_{\textrm{CH}_3} = -0.19$~eV
we can investigate resonance characteristics of the chemisorbed methyl group. Figure~\ref{fig:resonant_scatter} shows the change in DOS, $\Delta\nu(E)$, as well as, the resulting perturbed DOS per atom and spin, $\nu(E) = \nu_0(E)+\eta\Delta\nu(E)$, as functions of the Fermi energy for the admolecule
concentration $\eta = 0.001$\%.
The quantity $\nu_0(E) = -\frac{1}{\pi}\textrm{Im}\{G_0(E)\}=\tfrac{|E|}{D^2}$ is the DOS, per atom and spin, of the unperturbed graphene.
We clearly see a narrow peak at $E\simeq -8.8$~meV with a $\textrm{FWHM}\simeq 4.9$~meV in $\Delta\nu(E)$, indicating that $\textrm{CH}_3$ acts on graphene
as a strong resonant scatterer with the resonance close to the charge neutrality point.\\
\begin{figure}[htb]
\includegraphics[width=0.45\textwidth]{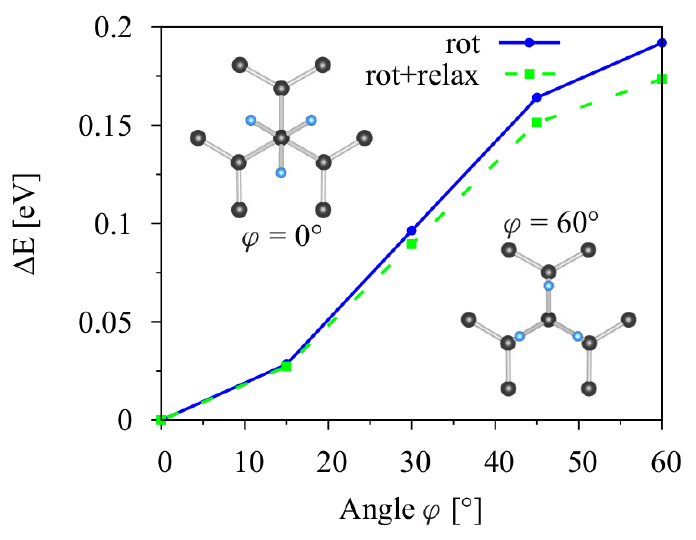}
\caption{\label{fig:angles}(Color online) Dependence of the total energy on the rotation angle $\varphi$ of the methyl group. The insets show the configuration of the admolecule in the cases $\varphi = 0^\circ$ and $ 60^\circ$. Label "rot" denotes the case where the admolecule was only rotated, label "rot+relax" denotes the case of rotation with a subsequent relaxation.}
\end{figure}
Figure \ref{fig:angles} shows the dependence of the total energy on the angle $\varphi$ of rotation of the methyl group with respect to the reference configuration corresponding to the fully relaxed 7$\times$7 supercell configuration used in all of our previous SOC calculations.
We performed two kinds of calculations. First, we rotated the admolecule around the C$_{\textrm{Ad}}$ - CH$_{3}$ bond ($z$-axis) without structural relaxation. Second, we also relaxed the structures at the given angle, keeping it fixed to avoid a back rotation during the relaxation process.
It turns out, that the rotation of the methyl group by an angle $\varphi = 60^\circ$ around the $z$-axis requires a maximum energy of about $0.17$~eV; see Fig.~\ref{fig:angles}. This corresponds to a temperature of $2090$~K or a frequency of $\omega = 41$~THz.
Comparing the maximum rotational energy with the thermal energy at room temperature $\textrm{k}_{\textrm{B}}\textrm{T} \approx 25$~meV, it is unlikely to rotate the methyl group at moderate temperatures by just thermal excitations. Another possibility would be terahertz radiation.
However, to couple the terahertz radiation to a molecule, a dipole moment oriented in a suitable direction needs to be present.
As a matter of fact, the CH$_3$ molecule in its pyramidal geometry possesses only an effective dipole moment along the $z$ axis. This is due to the different electronegativities~\cite{Allen1989} of hydrogen ($2.20$) and carbon ($2.55$), so the rotational excitation by terahertz radiation is not likely to be observed.\\
\section{\label{sec:Charge}Charge Density, Spin polarization and STM}
\begin{figure*}[htb]
\includegraphics[width=0.85\textwidth]{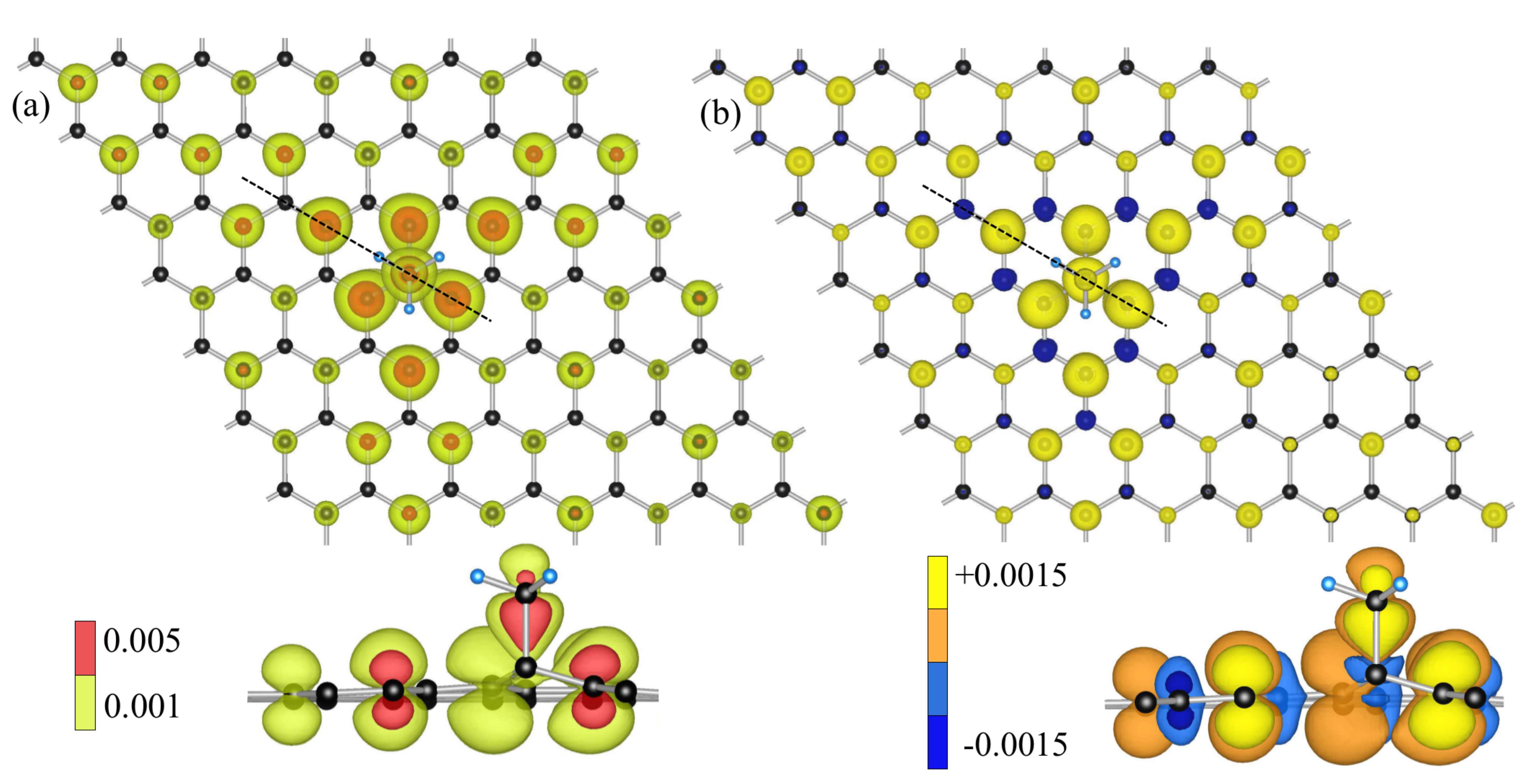}
\caption{\label{fig:spin_pol_ILDOS}(Color online) Methyl functionalized graphene in $7\times 7$ supercell configuration. (a) Top view of the electronic charge density. The charge density was obtained by summing the absolute squares of the Kohn-Sham states that lie in the energy between $-0.2$~eV and $0.2$~eV with respect to the Fermi level. (b) Top view of the spin-polarization. The spin-polarization was obtained by taking the difference between the spin-up and -down densities (see explanation in the text). Dashed lines in the top figures show directions for the cross-sectional views displayed at the bottom, the corresponding color values of the isosurfaces, in units (\AA$^{-3}$), are shown beside the cross-sectional views.}
\end{figure*}
In the left panel of Fig.~\ref{fig:spin_pol_ILDOS} we show the top view of the electronic charge density $\rho(\bm{r}) = \sum_{n,\bm{k}} |\phi_n^{\bm{k}}(\bm{r})|^2$ that is summed over the eigenstates $\phi_n^{\bm{k}}$ with energies $\varepsilon_n^{\bm{k}}$ in the energy window $\varepsilon_{\textrm{min}} = -0.2$~eV and $\varepsilon_{\textrm{max}} = 0.2$~eV with respect to the Fermi level.
The dashed line corresponds to the cross-sectional view displayed at the bottom. One sees a preferential localization of the electronic states mainly on the sublattice that is opposite to one where the methyl group is chemisorbed.
Therefore, the midgap band is formed mainly from states of the sublattice that contains C$_{\textrm{NN}}$ carbon atoms.
The charge density is strongly centered near the impurity, meaning the interaction among the different periodic images is negligible and a 7$\times$7 supercell is sufficient to represent the dilute limit.
It is worth to mention the pronounced triangular shape of the electronic charge density. Carbon atoms that mainly contribute to it are aligned along the directions spanned by the hydrogen atoms of the methyl group.
In general, this triangular anisotropy is universal for systems with $C_{3v}$ symmetry. It is also present in hydrogen~\cite{Sljivancanin2009} and fluorine~\cite{Irmer2014} functionalized graphene.\\
\begin{figure*}[htb]
\includegraphics[width=0.85\textwidth]{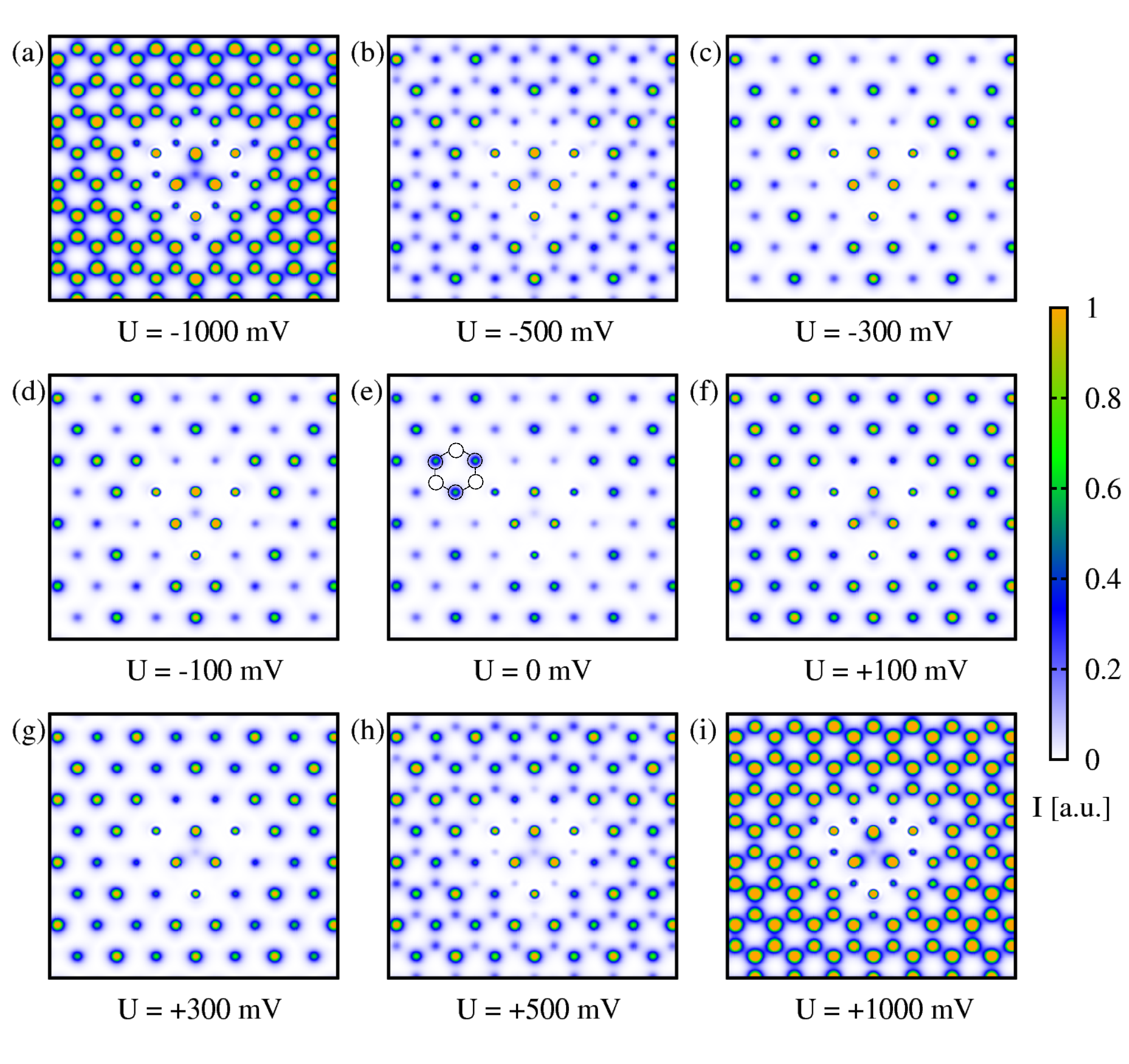}
\caption{\label{fig:STM}(Color online) Calculated STM images within Tersoff and Hamann approach~\cite{TersoffJ.andHamann1985}. States lying in the energy interval between $E_{\textrm{F}}$ and $E_{\textrm{F}} +eU$ are taken into account. Subfigures (a)-(i) correspond to biases between $-1$~V and $+1$~V. The color map gives the values of a tunneling current $I$, as defined in Eq. (\ref{eq:STM}), in arbitrary units. In subfigure (e) one graphene honeycomb is indicated as a guide for the eye.}
\end{figure*}
We also calculated the spin-polarization, $\Delta\rho= \rho_{\textrm{up}}-\rho_{\textrm{down}}$, for the 7$\times$7 supercell configuration; see right panel in Fig.~\ref{fig:spin_pol_ILDOS}.
The spin-polarization was obtained by taking the difference between spin up, $\rho_{\textrm{up}}$, and down, $\rho_{\textrm{down}}$, electronic densities. Each particular spin density
$\rho_{\textrm{up/down}}(\bm{r}) = \sum_{n,\bm{k}} |\phi^{\bm{k}}_{n,\textrm{up/down}}(\bm{r})|^2$ is obtained as a sum over the eigenstates with the corresponding spin polarization and energy that is below the Fermi level.
The dashed line in the top figure corresponds to the cross-sectional view displayed at the bottom. Each sublattice carries a different spin-polarization and
the \textit{up} contributions, are stronger than the \textit{down} ones.
This leads to a total magnetic moment of this structure of 1~$\mu_{B}$, which is in agreement with Ref.~\cite{Santos2012} and also in the line with Lieb's theorem, stating that an imbalance in the sublattice sites, leads to a net magnetic moment (see Ref.~\cite{PhysRevLett.62.1201}).
The local character of the spin-polarization, preferentially centered near the impurity, indicates that this is induced by the adsorbate.
The contributions from the hydrogen atoms of the methyl group are too small to be visible. Apart from the charge density, also the spin polarization
shares a clearly visible triangular shape anisotropy.\\
We also performed STM calculations, based on the formalism of Tersoff and Hamann~\cite{TersoffJ.andHamann1985}, as implemented in {\sc Quantum-ESPRESSO} code. The STM tunneling current is expressed as integral of the local density of states between the Fermi level $E_{\textrm{F}}$ and $E_{\textrm{F}}+eU$
\begin{equation}
I \propto \sum_{n,\bm{k}}|\phi^{\bm{k}}_{n}(\bm{r}_{0})|^2\Theta(E_{\textrm{F}}+eU-E_{n,\bm{k}})\Theta(E_{n,\bm{k}}-E_{\textrm{F}})
\label{eq:STM}
\end{equation}
with $\phi_{\nu}$ being states of the surface in the energy interval $E_{\textrm{F}}$ and $E_{\textrm{F}}+eU$ and $\bm{r}_{0}$ the position of the tip. The image provides information about occupied (unoccupied) states for negative (positive) biases $U$ that modifies the Fermi level.
Figure~\ref{fig:STM} shows calculated STM images for different bias voltages $U$. The fact that we also obtain an STM image for zero bias comes from a smearing contribution in energy, which is added to the bias $U$.
There are a few other features, which are noticeable. The first one is that for small biases (up to $\pm~300$~mV), only states of the sublattice opposite to methyl group contribute, as we already saw in Fig.~\ref{fig:spin_pol_ILDOS}.
Especially, this can be seen by looking at the graphene honeycomb, drawn for $U = 0$~V. For higher biases, we see that the other sublattice comes into play, but even at $\pm1$~V, some atoms in the vicinity of the admolecule are not as pronounced as others.
Another feature is that there is almost no difference between the images (c) and (d), even if the bias is changed by $200$~mV.
This is due to the fact that the DOS in Fig.~\ref{fig:bands_7x7} has two gaps at energies near the Fermi level and thus no additional states are available.
Next, we see that the images are symmetric with respect to the bias, which is not surprising, since the DOS shows this behavior, too. Finally, what we notice, is again the trigonal anisotropy; states sitting in the direction, where the hydrogen atoms point out, mainly contribute for small biases.

\section{\label{sec:Conclusion}Summary}
We have investigated SOC in graphene functionalized by the methyl group, a simple admolecule representing a wide class of organic compounds, analyzing
DFT-computed electronic band structures in the dense and dilute methyl coverage limits.
Compared to the pristine graphene, we have found a giant (100 times larger) SOC in the methyl functionalized graphene that originates from a local $sp^3$ distortion.
We have proposed a minimal realistic tight-binding model Hamiltonian and provided the relevant orbital and spin-orbital parameters that fit \textit{ab initio} computed band structure in the vicinity of the Fermi level.
As hydrogen, also the methyl group acts near the charge neutrality point of graphene as a narrow resonance scatterer. The minimal model Hamiltonian including the fitted tight-binding parameters can be used for further investigations of spin relaxation and spin transport, including the spin Hall effect characteristics that could be measured in graphene functionalized by light organic admolecules.
We have also analyzed conditions and energy ranges needed for excitations of rotational degrees of freedom of the methyl group.
Analyzing the calculated local densities of states, which simulate STM images, we found that the electronic density near the methyl admolecule shows a characteristic trigonal anisotropic shape, which could be potentially observed.
The magnitude of the induced SOC could be found directly from non-local spin Hall measurements on graphene with methyl adsorbates.

\section{\label{sec:Acknowledgments}Acknowledgments}
This work was supported by the DFG SFB 689 and GRK 1570, and by the European Union Seventh Framework Programme under Grant Agreement No.~604391 Graphene Flagship.
\bibliography{paper}

\end{document}